\documentclass[notitlepage,11pt,aps,pra]{revtex4-1}

\usepackage{graphicx}
\usepackage{subfigure}
\newfont{\rsfsten}{rsfs10 scaled 1200}
\newfont{\rsfsseven}{rsfs10 scaled 1200}
\newfont{\rsfsfive}{rsfs10 scaled 1200}
\usepackage{color}
\usepackage{amsmath}
\usepackage[utf8]{inputenc}
\usepackage[margin=0.95in]{geometry}

\newcommand\norm[1]{\left\lVert#1\right\rVert}

\begin{document}

	\title{Integration of Few Body Celestial Systems Implementing Explicit Numerical Methods}
	
	\author{Achilleas Mavrakis}
	\affiliation{National Technical University of Athens, Zografou, Athens, 15780, Greece}
	\author{Konstantinos Kritos}
	\affiliation{National Technical University of Athens, Zografou, Athens, 15780, Greece}
	
	\date{\today}
	
	\begin{abstract}
	
	The $N$-body problem is of historical significance because it was the first implementation of the Newtonian dynamical laws for the description of our Solar System.
	Motivated by this, the project's goal is to revisit this problem for small $N$ and find a solution for the trajectories of specific two-body and three-body configurations as well as the planetary orbits of our Solar System using a fourth order Runge-Kutta explicit iterative method. We find an adequate agreement in our results with planetary trajectories found online.
	
	\end{abstract}
	
	\maketitle
	
	\section{Introduction}
	
	The two-body problem is one of the most well-known problems in the literature and can be found in most classical mechanics textbooks and reviews, for instance in \cite{collins,arnold,mayer,landau}. Analytical methods can be invoked for the solution to the two-body problem and be given in a closed form. In the three-body problem, there exists a slowly converging power-series solution only under certain conditions as was demonstrated early on by \cite{sundman,poincare}. However, under a more general setting, the solution to the three-body, let alone the many body problem, can be addressed numerically and simple methods can be invoked to solve the few body problem \cite{anagnostopoulos,ydri}.
	
	In the past there have been various numerical \cite{aarseth,szebehely,bettis,nacozy,ahmaad,zadunoisky,mikkola} and semi-analytical \cite{musielak,naoz,laves,becker,arenstorft,wang} surveys on the $N$-body problem and its applications \cite{hayli, wielen}. Stability \cite{szebehelyStab,scheeres,marchal,elmabsoutStab} and error analysis studies \cite{baba,dejonghe} was also investigated. The study of the $N$-body problem has become popular especially after the advent of modern technology and computer power, resulting in the development of advanced algorithmic techniques and parallel computing \cite{Hertz,belleman}.
	Regularization schemes have also been developed to overcome singularities encountered in numerical simulations \cite{szszodr,heggie,zare,MikkolaReg}.
	For a comprehensive book on the study of N-body systems and their numerical simulations see \cite{aarsethBOOK}.
	
	In this project, we aim at integrating the few body problem implementing low order explicit numerical methods. Specifically, we define the $N$-body problem in three dimensions and apply for small $N$ (here at most $N=9$). We consider two-body and three-body bounded systems like the Pluto-Charon pair and a star-planet-comet triple. Finally, we simulate the Solar System taking into account only the eight planets and the Sun.
	We test whether the well-known 4-stage explicit Runge-Kutta iterative method (RK4) \cite{butcherBook, butcherRunge}, which is of relatively low computational expense and of non-significant stability, can be justified and sufficiently approximate this problem, instead of applying more heavy computational or greater stability-wise algorithms such as implicit, Predictor-Corrector (PECE) \cite{freed}, and semi-unconditional methods \cite{butcherImplicit, butcherStabImplicit,gautchi}.
	Here, we have developed our own numerical codes writing the RK4 algorithm routine from scratch in FORTRAN language.
	
	The rest of this work is developed as follows. In Sec.~\ref{method_sec} we define the N-body problem and the numerical method we implement and in Sec.~\ref{result_sec} we present the results of the integration we perform. Finally, in Sec.~\ref{conclusions_sec} we give our discussion and conclude.

	\section{Method}
	\label{method_sec}
	
	\subsection{The N-body gravitational problem}
	An $N$-body configuration, composed of $N$ point masses, is determined by the time-dependent positions, $q^i_\alpha(t)$, and the momenta, $p_{i\hspace{0.3mm} \alpha}(t)$, for each member of the system. Here, $t$ is the time, $i=1,2,3$ denotes the components and $\alpha=1,...,N$ labels each particle. Such a configuration whose members interact solely via gravitational forces, is defined as a Hamiltonian system of the from 
\begin{subequations}
\begin{align}
	\dot{\boldsymbol{q}}_\alpha&=\dfrac{\boldsymbol{p}_\alpha}{m_\alpha},\\
	\dot{\boldsymbol{p}}_\alpha&=-\sum_{\beta=1,\ \beta\ne\alpha}^N Gm_\alpha m_\beta\dfrac{\boldsymbol{q}_\alpha-\boldsymbol{q}_\beta}{\norm{\boldsymbol{q}_\alpha-\boldsymbol{q}_\beta}^3},
\end{align}
\label{Nbody_eqs}
\end{subequations}
	where $G$ is the gravitational constant, which we take to be unity.	
	This system is conservative and theoretically speaking the total energy, E, and total angular momentum, L, are conserved and constant in time. They are constants of motion. Monitoring the time evolution of the values of E and L given a particular system throughout a simulation, is of importance. Because, if E and L are not constant in time, this indicates that the simulation failed. Small deviations are expected, as long as they are in agreement with the error of the respective numerical method. Finally, the $N$-body configuration is a ``stiff'' problem, thus not all numerical methods give reliable solutions and we would need a special type of stability. (A differential system is said to be stiff, when the Jacobian has at least a large negative eigenvalue.)

	\subsection{The numerical algorithm}
	\label{num_sec}
	The $N$-body problem is quite complicated for $N\ge3$ it cannot always be solved analytically and a complete analytic solution exists only for $N=2$. For this reason, we resort to numerical solutions. In this project, we are going to use the 4-stage Runge-Kutta explicit iterative method (RK4) to solve the few body problem ($N\leq 9$) in three dimensions \cite{butcherBook,butcherRunge,gautchi}. As it will be noted, this is not the best fitted numerical solution for this problem, mainly if one is interested in long running simulations. Due to the fact that it is explicit and not an (semi-)unconditional method, it is not suitable for stiff problems. We would need an unacceptably small step size level in order to better approximate the theoretical solution, which would make the process quite slow. So, for this reason we will mainly use a relatively small step size. But it is intriguing nonetheless to see how a method, with relatively low computational expense and no noteworthy stability, fairs with such a problem, as an alternative for methods with heavy computational needs and greater stability (e.g. A-stability).

	As we can see from Appendix~\ref{RK_appendix}, by choosing the node and weight parameters $\tau_i, w_i, a_{ij}$, we can produce Runge-Kutta methods \cite{butcherCoeff} that are consistent and have sufficient stability in order to achieve the maximum possible order. Therefore, choosing the parameters in the form of Matrices as below,
\begin{align}	
	\begin{array}{c|c}
		0 \  0 \ 0 \ 0 & 0 \\ 
		{\frac{1}{2}} \ 0 \  0 \ 0  & {\frac{1}{2}} \\
		0 \ {\frac{1}{2}}\  0 \  0  & \frac{1}{2} \\
		0 \ 0 \  1 \  0  & 1 \\
		\hline
		\frac{1}{6} \ \frac{1}{3} \  \frac{1}{3} \ \frac{1}{6}
	\end{array}
\end{align}
we get the 4-stage explicit Runge-Kutta formulae. Best written as,
\begin{align}
		\label{106}
		\left\{\begin{array}{ll}{k^{n,1}=F(t^n,z^n)} & \mbox{} \\
		{k^{n,2}=F(t^n + \frac{1}{2}h,z^n +\frac{1}{2}hk^{n,1})} & \mbox{} \\
		{k^{n,3}=F(t^n +\frac{1}{2}h,z^n+ \frac{1}{2}hk^{n,2})} & \mbox{} \\
		{k^{n,4}=F(t^n +h,z^n +hk^{n,3})} & \mbox{} \\
		{z^{n+1}=z^n + \frac{h}{6}(k^{n,1}+2k^{n,2}+2k^{n,3}+k^{n,4})} & \mbox{}
		\end{array}\right\}
\end{align}
	
It has been proven by M.~W.~Kutta in the 1900s, that such explicit methods of stage $1 \leq p \leq 4$, attain maximum order of exactly $p$. That means that, our method has global order of 4 ($\mathcal{O}(h^4)$), since we used 4-stages. This can be proven by application of Taylor's expansion for (vector-valued) functions, though for high orders like our method it is technically much more demanding than the lower ones. Therefore, one can use graph-theoretical tools, as established by J.C. Butcher \cite{butcherRungeTree}.

	\section{Results}
	\label{result_sec}
	In the previous sections we defined the N-body problem, as well as the numerical integration algorithm that we are going to use. In this section, we are going to present numerical solutions to three celestial systems for $N=2$ bodies (Pluto-Charon couple), $N=3$ bodies (star-planet-comet triple) and $N=9$ bodies (our Solar System). The correct choice of initial conditions is of importance for all the following examples, so that we have a bounded and periodic movement in which there are no collisions. In case the bodies get too close and collide, the gravitational force will become huge and the equation will diverge to infinity. This is a type of singularity, as the position vectors are inversely proportional to the square of the distances between two interacting bodies. Reminder, for all of the solutions that follow below, we have worked in a unit system in which the gravitational constant is equal to one ($G=1$).

	\subsection{Pluto-Charon}
\begin{figure}[h]
	\centering
	\includegraphics[width=0.7\textwidth]{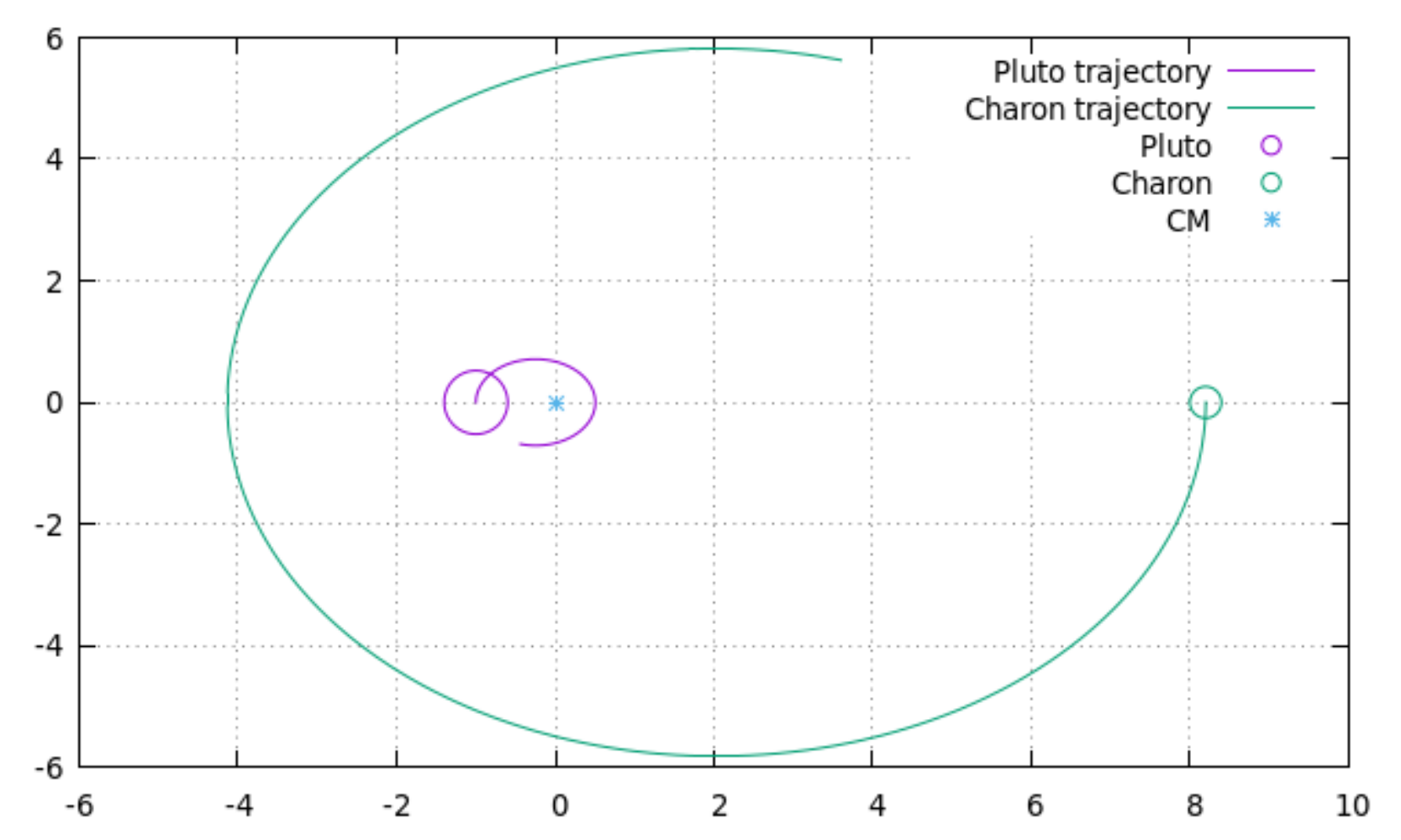}
	\caption{Pluto-Charon System.}
	\label{Pluto_fig}
\end{figure}

	In this example, we are going to simulate the celestial system of Pluto and its moon Charon, neglecting the gravitational effects from other moons, other celestial objects and planets as they are irrelevant. It is a unique pair, since Charon is such a large moon compared to its parent body that the barycenter of the system lies outside of Pluto which is the primary mass. This means that Pluto orbits about a point which lies at a distance larger than its radius. Charon orbits in an opposite fashion to Pluto about the center of mass, always respecting the conservation laws. From data known, the mass of Charon is $12.2\%$ the mass of Pluto (\cite{null}, \href{https://solarsystem.nasa.gov/moons/pluto-moons/charon/by-the-numbers/}{https://solarsystem.nasa.gov/moons/pluto-moons/charon/by-the-numbers/}). As for the initial conditions we take those seen in Table~\ref{plutoCharonTable} and assume that the center of mass is at the point $(0,0,0)$. Pluto's initial position is taken to be at the arbitrary $(-1,0,0)$ and by use of momentum conservation, we determine the initial position and velocity of Charon. We present part of the trajectory (about three quarters to a complete period) in Fig.~\ref{Pluto_fig}.

\begin{figure}[h]
	\centering
	\includegraphics[width=0.7\textwidth]{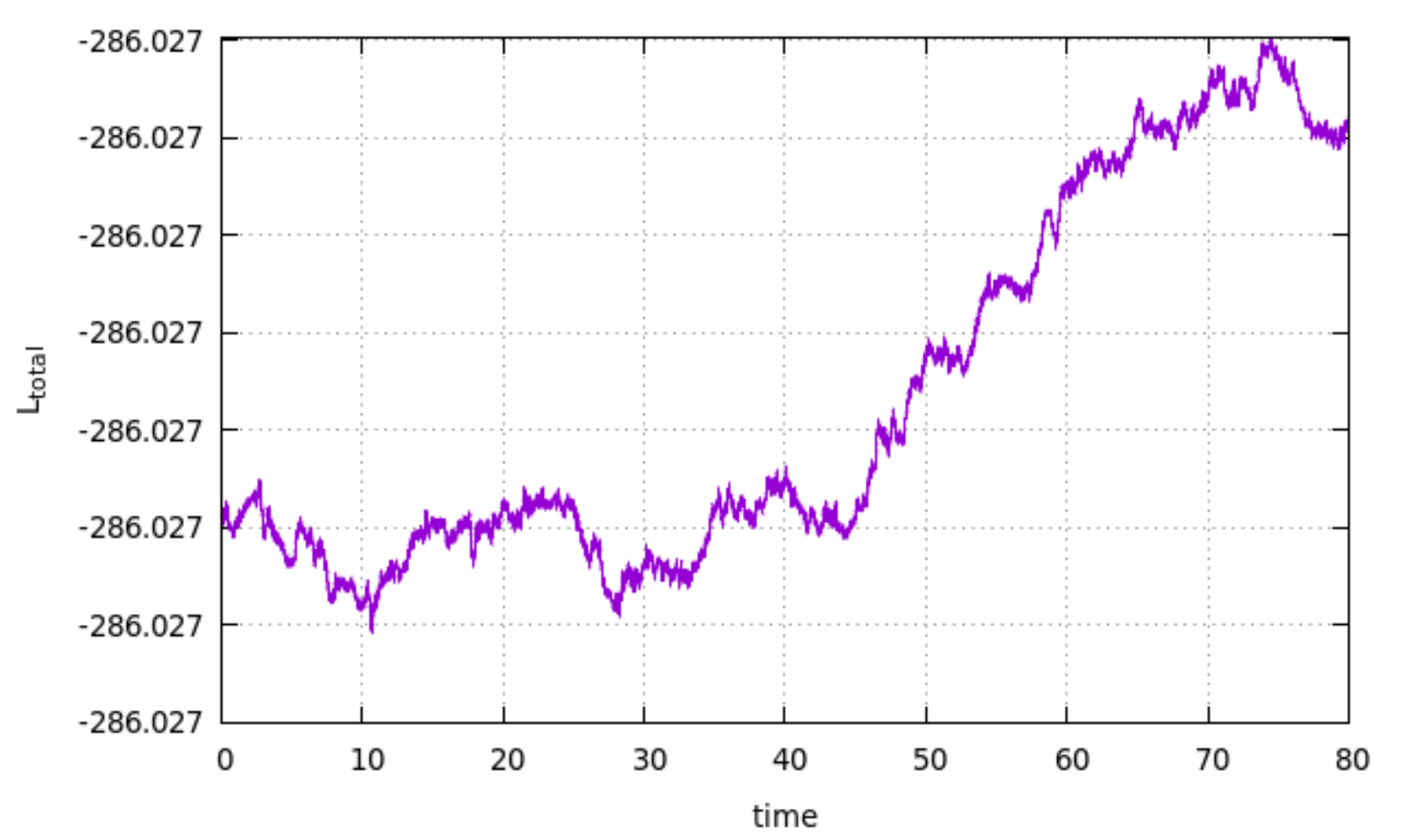}
	\caption{Angular Momentum of Pluto-Charon.}
	\label{LPluto_fig}
\end{figure}

\begin{figure}[h]
	\centering
	\includegraphics[width=0.7\textwidth]{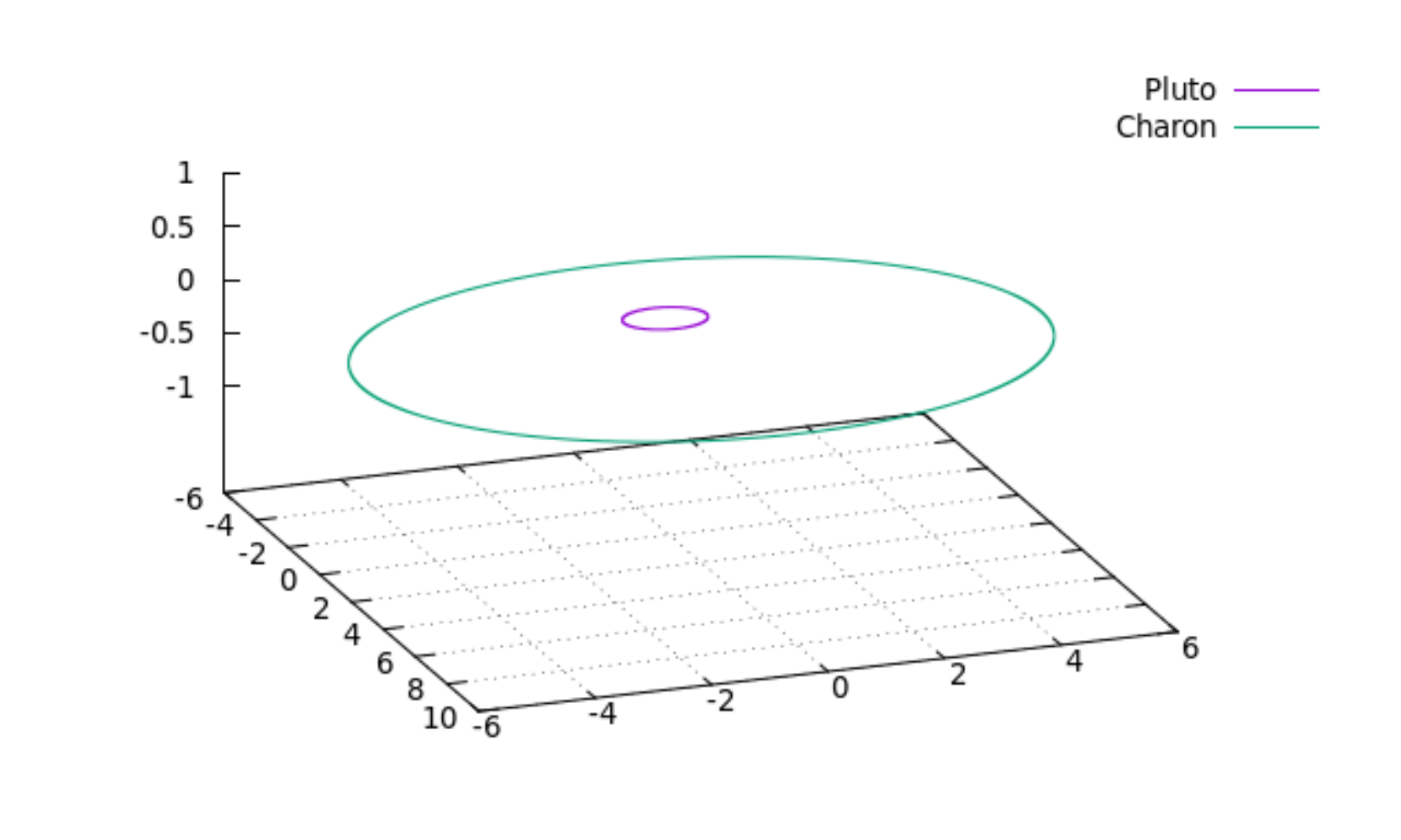}
	\caption{Plane Movement of Pluto-Charon}
	\label{Pluto3d_fig}
\end{figure}

	Since the total angular momentum of a bounded system of $N$-bodies is an integral of motion, we get that the third component of the total angular momentum equals
	$$L_3=q_1^1p_{12}-q_1^2p_{11}+q_2^1p_{22}-q_2^2p_{21}=-286.027$$ 
	with $L_1=L_2=0$. We notice that it is constant and independent of time. As we can also see from Fig.~\ref{LPluto_fig}, this conservation is respected in our simulation, while small changes in values are of order $10^{-11}$, which is in agreement with the error predicted by the RK4 method. The conservation of the angular momentum vector is also consistent with the plane movement seen in Fig.~\ref{Pluto3d_fig}, as it is orthogonal to the subspace created by the vectors $\boldsymbol{q}$ and $\boldsymbol{p}$. Further, the energy of the system has also been checked to be constant within the errors of the method used.

	\subsection{Star-Planet-Comet}
	
	Often, in $N$-body systems, one seeks for a simplistic solution of the problem for a qualitative description of the orbits. As an example, for a more precise calculation of the Earth's orbit, it would suffice to take into consideration only the Sun and the large planets while we may ignore the gravitational effect of smaller celestial objects, like moons, dwarf planets, asteroids and comets. Motivated by such an approximation, in this sub-section we consider a system of three masses with a hierarchy in their values. This is often referred to as the restricted three-body problem \cite{pojoman}. In particular, we consider a central heavy object representing a star, along with a light planet orbiting around that star and a third even lighter body, which could represent a comet. So, we have the case $m_\text{star}\gg m_\text{planet}\gg m_\text{comet}$. To find an approximate numerical solution to the differential system in Eq.~\eqref{Nbody_eqs} we utilize this mass hierarchy to our benefit. For instance, to calculate the force acted on the planet or the comet we only take into consideration the contribution from the star while the effect of the planet on the comet is a small perturbation which we can be neglected. Such a planet-comet interaction would only affect the orbit of the comet in case the latter encounters the planet closely, in which case the forces would be larger. The corresponding differential system is given by,
	\begin{subequations}
		\begin{align}
			\ddot{\boldsymbol{q}}_\text{\tiny star}&\simeq0,\\
			\ddot{\boldsymbol{q}}_\text{\tiny planet}&\simeq-Gm_\text{\tiny star}\frac{\boldsymbol{q}_\text{\tiny planet}-\boldsymbol{q}_\text{\tiny star}}{\norm{\boldsymbol{q}_\text{\tiny planet}-\boldsymbol{q}_\text{\tiny star}}^3}+\underbrace{{\cal O}(m_\text{\tiny comet})}_\text{\tiny we neglect},\label{planetAccel}\\
			\ddot{\boldsymbol{q}}_\text{\tiny comet}&\simeq-Gm_\text{\tiny star}\frac{\boldsymbol{q}_\text{\tiny comet}-\boldsymbol{q}_\text{\tiny star}}{\norm{\boldsymbol{q}_\text{\tiny comet}-\boldsymbol{q}_\text{\tiny star}}^3}\underbrace{-Gm_\text{\tiny planet}\frac{\boldsymbol{q}_\text{\tiny comet}-\boldsymbol{q}_\text{\tiny planet}}{\norm{\boldsymbol{q}_\text{\tiny comet}-\boldsymbol{q}_\text{\tiny planet}}^3}}_\text{\tiny relevant only during a strong planet-comet encounter}.\label{cometAccel}
		\end{align}
	\label{starPlanetCometEq}
	\end{subequations}
	The condition for which a planet-comet interaction is not negligible is when the force of the planet on the comet is of the same order as the force of the star on the same comet; i.e. $m_\text{star}\norm{\boldsymbol{q}_\text{\tiny comet}-\boldsymbol{q}_\text{\tiny planet}}^2\sim m_\text{\tiny planet}\norm{\boldsymbol{q}_\text{\tiny comet}-\boldsymbol{q}_\text{\tiny star}}^2$. However, since $m_\text{\tiny comet}\ll m_\text{\tiny planet}$, the deviation of the planet's trajectory due to a strong planet-comet interaction will not be affected significantly, and thus we neglect the force of the comet on the planet (the second term in the right-hand-side of Eq.~\eqref{planetAccel}). Nevertheless, in the time interval for which we have integrated Eq.~\eqref{starPlanetCometEq} there are no strong encounters between the planet and the comet and therefore the second term in the right-hand-side of Eq.~\eqref{cometAccel} is irrelevant in our case. Initial conditions for Eq.~\eqref{starPlanetCometEq} are shown in Table~\ref{case2}.
			
	\begin{figure}
		\centering
		\includegraphics[width=1\textwidth]{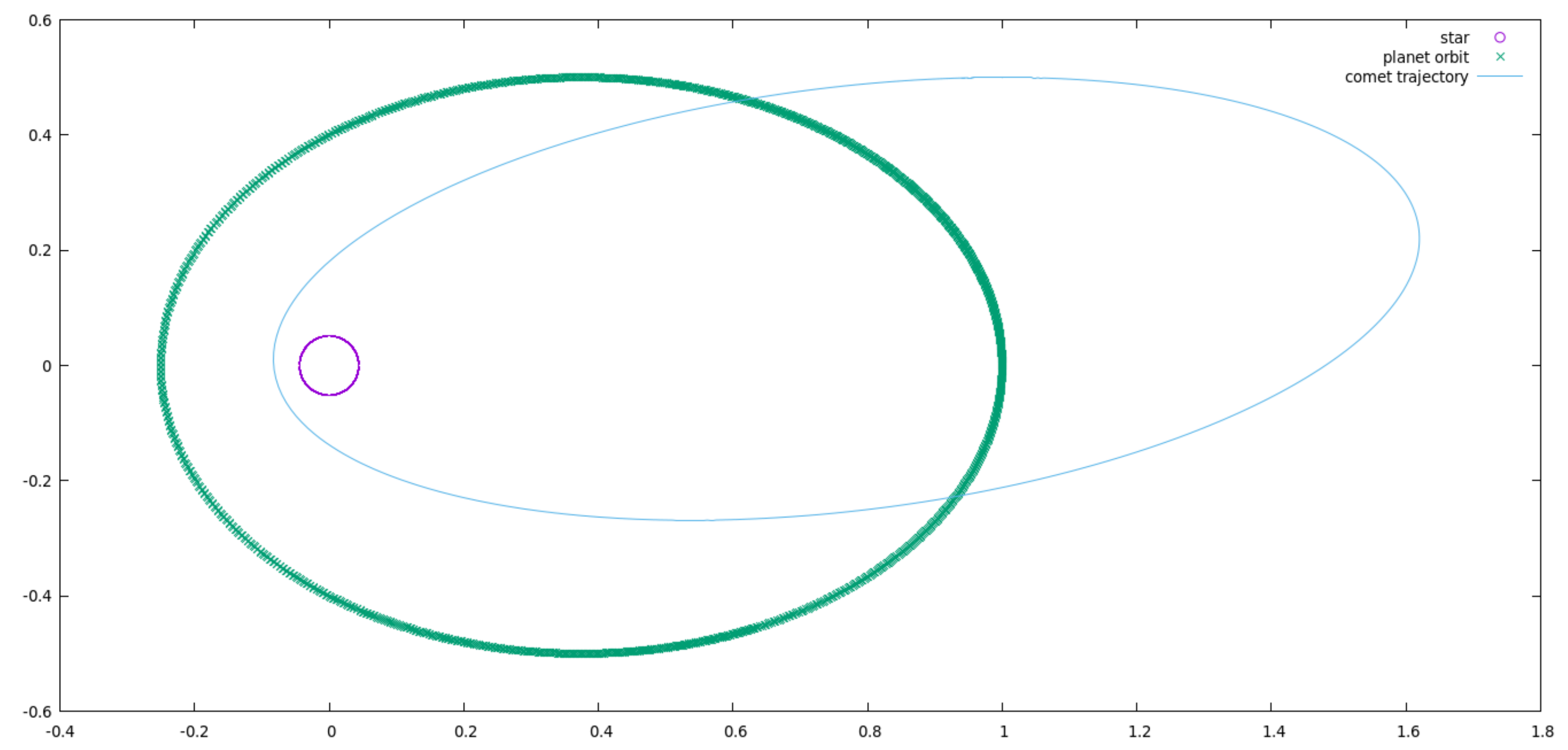}
		\caption{Trajectories in the Star-Planet-Comet simulation. The star is represented by the purple circle which is not a trajectory.}
	\end{figure}

	\subsection{Solar System}
\begin{figure}[t]
		\centering
		\includegraphics[width=1\textwidth]{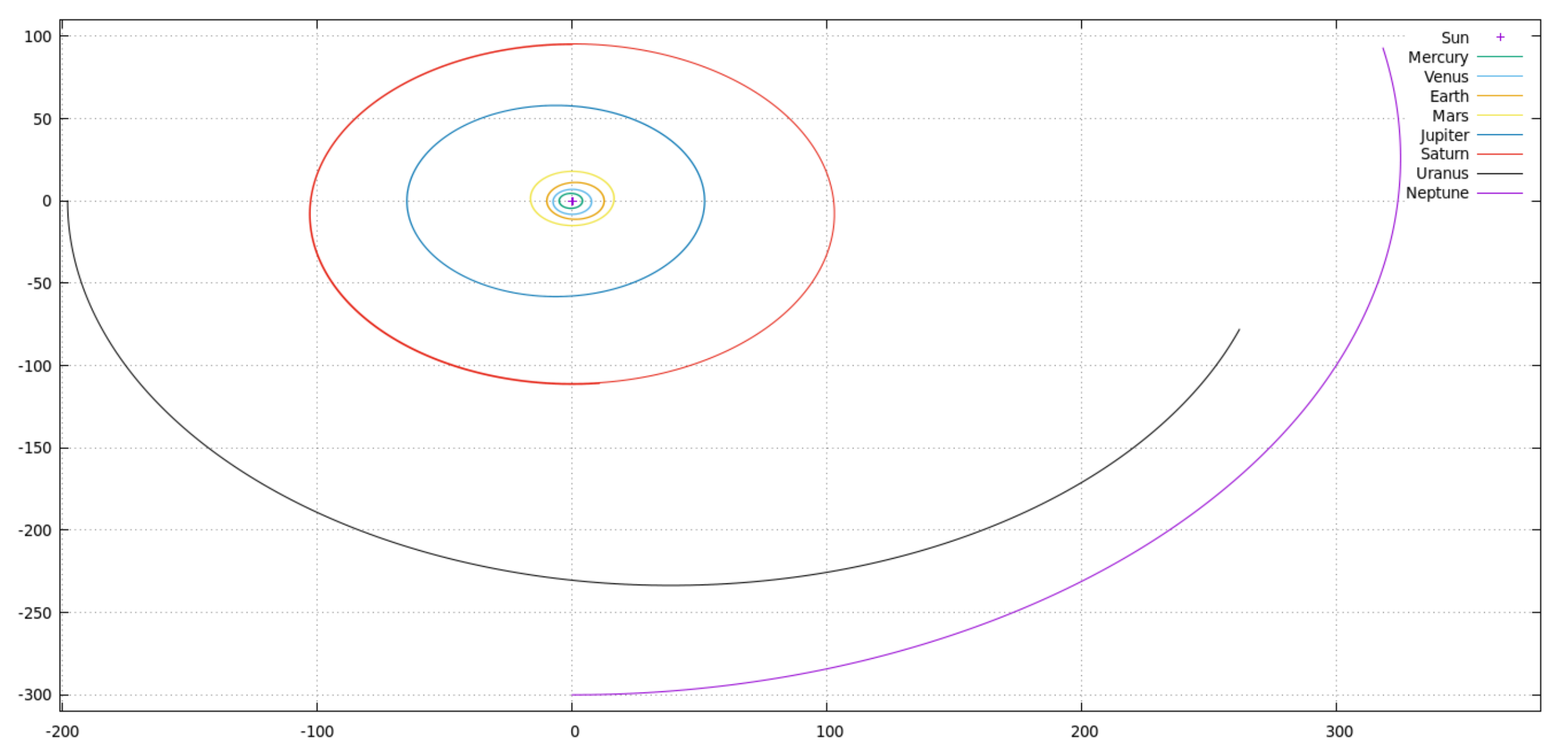}
		\caption{Solar System Simulation}
		\label{solarsystem_fig}
	\end{figure}

	In this last example, we attempt to calculate the orbits of the planets in our Solar System. The Sun comprises nearly all of the matter in the Solar System, reaching up to $99.86\%$ of total mass and is expected to have a nearly zero acceleration in the Solar System frame, \href{https://solarsystem.nasa.gov/solar-system/sun/overview/}{https://solarsystem.nasa.gov/solar-system/sun/overview/}. According to Kepler's laws of planetary motion, each object will travel along an ellipse with the Sun being approximately stationary at one of the focal points. Consequently, the closer a planet is to the Sun, the faster it will revolve around it and the smaller its orbital period. Therefore, by the time an outer planet (e.g. Jupiter) completes a full revolution, an inner planet (like Venus) will have already completed several orbits. In particular, in our simulations, we have integrated the system over such a time interval such that Uranus and Neptune orbits have not closed (see the black and purple lines in Fig.~\ref{solarsystem_fig}).

	We treat the Solar System as a $N=9$ body problem, where the initial conditions and masses of the planets have all been normalized in terms of the Earth's. We consider a system of units in which the mass of the Earth is taken to be $m_\text{\tiny Earth}=1$.
	The mass of the Sun and of the other planets are then computed according to the relative mass ratio normalized to $m_\text{\tiny Earth}$, \href{https://solarsystem.nasa.gov/planets/overview/}{https://solarsystem.nasa.gov/planets/overview/}. The initial conditions are given in Table~\ref{case3}, where the conversion factor of $0.03$ for the mass has been used so that the ratio between distances and masses is realistic in our numerical model. In order to conserve momentum at the center of mass frame (which almost coincides with the position of the Sun), we take the initial positions and momenta of the planets to cancel out. Finally, we present the trajectories obtained from the numerical calculation in Fig.~\ref{solarsystem_fig}.

\begin{figure}[t]
		\centering
		\includegraphics[width=0.7\textwidth]{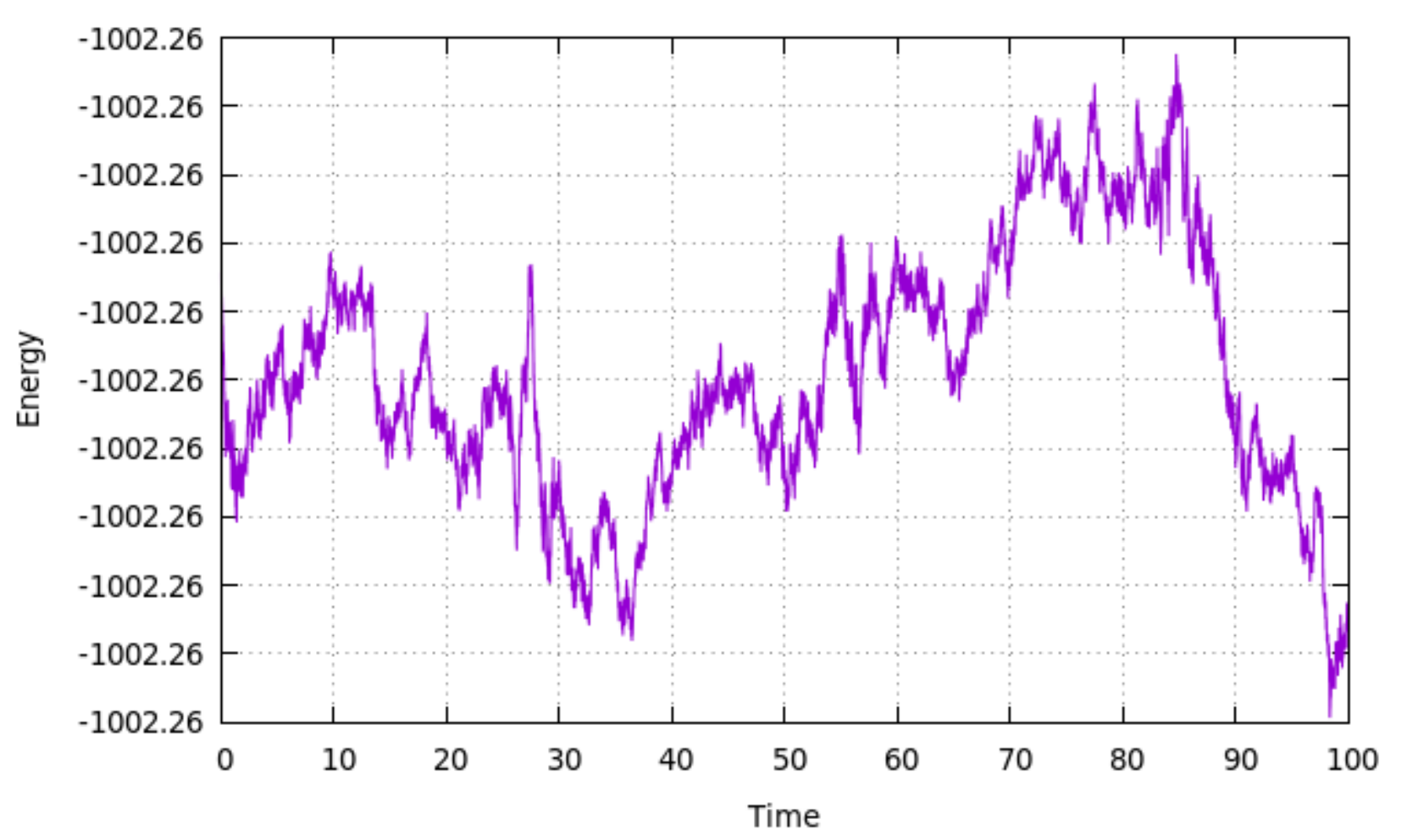}
		\caption{The time evolution of the energy in the Solar System example.}
		\label{Esolarsystem_fig}
	\end{figure}

	We also give the energy over time in Fig.~\ref{Esolarsystem_fig}, where we notice that the fluctuations displayed are within the error margin of the 4-stage Runge-Kutta method. Nevertheless, it is approximately constant as expected from a theoretical standpoint. The angular momentum was also monitored and found to be conserved within errors and fluctuates like the energy in a similar manner, but it is not depicted here.
	    	
\begin{figure}[h]
		\centering
		\includegraphics[width=0.7\textwidth]{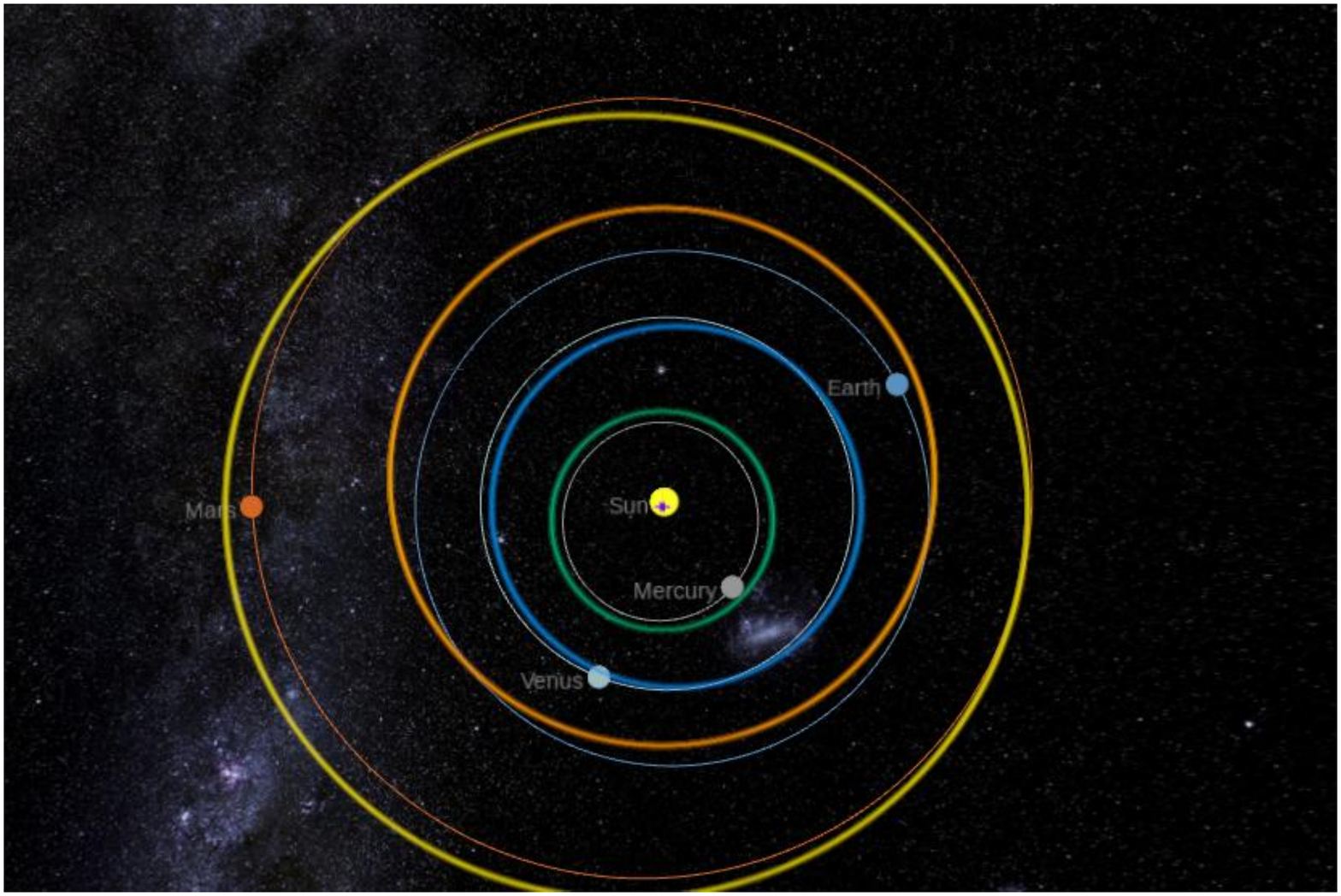}
		\caption{The numerical trajectories (solid orbits) of the inner planets in comparison with the corresponding trajectories (faint orbits) as obtained from \href{https://theskylive.com/3dsolarsystem}{https://theskylive.com/3dsolarsystem} .}
		\label{compare_fig}
	\end{figure}

	Moreover, in Fig~\ref{compare_fig} we compare the orbits of the inner planets (Mercury, Venus, Earth and Mars) obtained from our numerical model with orbits found online (\href{https://theskylive.com/3dsolarsystem}{https://theskylive.com/3dsolarsystem}). Looking at Fig.~\ref{compare_fig}, we notice that there are the small differences in the trajectories of the planets. More generally, planets nearest to the Sun do not stray as much as planets that are further away (like Uranus, which is not depicted in Fig.~\ref{compare_fig}) for which we find that the difference in their orbits is larger. Nevertheless, our trajectories were calculated using a computational inexpensive scheme with relatively small step size. Since, as we already said the initial conditions affect our solution curves, they also play a significant role in the deviation of the two sets of orbits.
	For that reason it was relatively a successful simulation. For better results, a smaller step size would probably not suffice in the long run and a more sophisticated analysis would be needed (see Sec.~\ref{conclusions_sec}).
		
	\section{Discussion}
	\label{conclusions_sec}
	
	To summarize, in this project we implemented the RK4 method in the context of few-body celestial systems and observed the solution curves and the behavior of the bodies in those systems. We aimed to simulate these models by using a relatively low computational cost numerical method with fast results and easy programming. So we settled with the explicit 4-stage Runge-Kutta method and saw whether it could be efficient enough to solve stiff problems, with the $N$-body problem being one them. Eventually, we noticed that for celestial systems with fewer interaction terms (e.g. Pluto-Charon system), that is with $N$ small, this method had sufficient and quick results as seen in the Sec.~\ref{result_sec}. However, since the equations are stiff, after a certain number of periods, the orbits eventually diverge, unless we use some step-size control scheme, where we would only delay this effect and make the whole process too slow. This is the biggest downside of this numerical method, as it lacks in stability, being a 0-stable and not unconditionally stable, which is needed in our case. Albeit the fact that with few interactions the results where ample, when more celestial bodies came into play, the stiffness became more predominant altering the orbits (e.g. in the Solar System example) before finishing a full period. Although, the simulation, as already mentioned was somewhat successful, one cannot but notice the difference in the trajectory of the planets with the orbits found elsewhere (see Fig.~\ref{compare_fig}). Making the step-size smaller with some step-size control schemes, seemed to fix the problem for the first few periods only, but it needed way too much time to finish, which is something we wanted to avoid from the beginning. So, we came to the conclusion that a more efficient way to solve this problem, was to not simulate it with the RK4 method, as it was too unreliable, as expected, for that many bodies. A general method that could be used instead, is an implicit Runge-Kutta method, but the considerable computational expense involved with it cannot be justified for this problem. More fitted well-known numerical schemes would be the semi-implicit Runge-kutta or PECE methods, which, again, are more expensive than a regular RK4 method. Since, for the former in each step we would need to solve a system of equations, and for the latter one, twice the evaluation of the function $F$ per step would be required. An even better treatment of this problem, would be to implement a semi-unconditional method, like the non-trivial backward differentiation formula methods (BDF-s for $s>1$).
	In conclusion, the results of this project showed that, for few body systems and for few periods, the explicit RK4 is sufficient, but when the problem has more interactions, then we believe that an unconditional or semi-unconditional method is required for its simulation.
	\\	

	This work was developed in the summer of 2019 and was translated into English by the authors in the winter of 2020.

\begin{table}[h]
	\centering
	\begin{tabular}{|c|c|c c c|c c c|}
		\hline
		Object & mass & $q^{1}$ & $q^{2}$ & $q^{3}$ & $p_{1}$ & $p_{2}$ & $p_{2}$ \\
		\hline
		Pluto	& 100 & -1 & 0 & 0 & 0 & 31.1 & 0 \\
		Charon	& 12 & 8.197 & 0 & 0 & 0 & -31.1 & 0 \\
		\hline
	\end{tabular}
	\caption{Initial conditions for the Pluto-Charon system.}
	\label{plutoCharonTable}
\end{table}

\begin{table}[h]
	\centering
	\begin{tabular}{|c|c|c c c|c c c|}
		\hline
		Object & mass & $q^{1}$ & $q^{2}$ & $q^{3}$ & $p_{1}$ & $p_{2}$ & $p_{2}$ \\
		\hline
		Star	& $10^5$ & 0 & 0 & 0 & 0 & 0 & 0 \\
		Planet	& 1 & 1 & 0 & 0 & 0 & $2\times10^2$ & 0 \\
		Comet	& $10^{-5}$ & 1 & 0.5 & 0 & -2.5 & -3 & 0 \\
		\hline
	\end{tabular}
	\caption{Initial conditions for the Star-Planet-Comet example.}
	\label{case2}
\end{table}

\begin{table}[h]
	\centering
	\begin{tabular}{|c|c|c c c|c c c|}
		\hline
		Object & mass$/0.03$ & $q^{1}$ & $q^{2}$ & $q^{3}$ & $p_{1}$ & $p_{2}$ & $p_{2}$ \\
		\hline
		Sun		& 333043.1 & 0  & 0 & 0 & 28.2 & -134.94 & 0 \\
		Mercury	& 0.055 & 4 & 0 & 0 & 0 & 0.0875 & 0 \\
		Venus	& 0.815 & 0 & 7 & 0 & -0.95844 & 0 & 0 \\
		Earth	& 1 & -10 & 0 & 0 & 0 & -1 & 0 \\
		Mars	& 0.1075 & 0 & -15 & 0 & 0.08686 & 0 & 0 \\
		Jupiter	& 317.82 & 52 & 0 & 0 & 0 & 139.2 & 0 \\
		Saturn	& 95.081 & 0 & 95 & 0 & -30.42 & 0 & 0 \\
		Uranus	& 14.54 & -198 & 0 & 0 & 0 & -3.3442 & 0 \\
		Neptune	& 17.148 & 0 & -300 & 0 & 3.087 & 0 & 0 \\
		
		\hline
	\end{tabular}
	\caption{Initial conditions for the Solar System. The masses and momenta are normalized with respect to the Earth.}
	\label{case3}
\end{table}

	\appendix
		
	\section{Formulation of Runge-Kutta types}
	\label{RK_appendix}
	In this appendix we formulate the general Runge-Kutta method for random coefficients considering an ordinary differential equation (ODE) of the form :
	\begin{align}	
			\label{101}
			\left\{\begin{array}{ll}{z'=F(t,z)}, & \mbox{ } t \in [a,b], \\{z(a)=z_0. } & \mbox{ } 
			\end{array}\right.		    
	\end{align}

We begin by applying a uniform discretization  at the interval $[a,b]$. So if we separate the interval into $N$ sub-intervals, we get that the discretized time-step is $t^n=a + nh$, where $h=\dfrac{b-a}{N}$. Then by integrating Eq.~\eqref{101} from $t^n$ to $t^{n+1}$ we get :
  \begin{align}
  		z(t^{n+1})=z(t^n) + \int_{t^n}^{t^{n+1}}F(t,z(t))dt.   
  \end{align}	
Then considering the approximations $z(t^n)\approx z^n$,$z(t^{n+1})\approx z^{n+1}$ and the change in coordinates  $t=t^n +hs$, we conclude that 
\begin{align}
	z^{n+1}=z^n + h\int_{0}^{1}F(t^n+hs,z(t^n +hs))ds. 
\end{align}
Then according to Gauss-Legendre numerical integration with $q$ nodes $\tau_i$, $i=1,..,q$ on the interval $[0,1]$ and the corresponding weights $w_i$ , it can be rewritten as 
\begin{align}
		\label{103}
		z^{n+1}=z^n + h\sum\limits_{i=1}^{q}w_iF(t^n +h\tau_i,z(t^n +h\tau_i)),
\end{align}
where $\tau_i$ and $w_i$ are chosen accordingly. In order to calculate the remaining term $z(t^n +h\tau_i)$ we once again integrate Eq.~\eqref{101}, this time from $t^n$ to $t^{n,i}:=t^n +h\tau_i$ with the same change in coordinates, extracting the following set of equations 
\begin{align}	
		z(t^{n,i})=z(t^n) + h\int_{0}^{\tau_i}F(t^n+hs,z(t^n +hs))ds, \ \ i=1,...,q.
\end{align}
Repeating the Gauss-Legendre numerical integration for every node $\tau_i$, $i=1,..,q$ and considering the approximations $z(t^{n,i})\approx z^{n,i}$,$z(t^{n})\approx z^{n}$ we get 
\begin{align}
		\label{105}
		z^{n,i} = z^{n} +h\sum\limits_{j=1}^{q}a_{ij}F(t^{n,j},z^{n,j}), \ \ i=1,...,q.
\end{align}
where the coefficients $a_{ij}$ are the corresponding weights at every node $\tau_i$, $i=1,..,q$. So the above relation represents a non-linear system of $q$ equations with $q$ unknown parameters.
Summarizing, the method can be written as 
\begin{align}
		\label{104}
		\left\{\begin{array}{ll}{z^0=z_0} & \mbox{}  \\{ z^{n,i} = z^{n} +h\sum\limits_{j=1}^{q}a_{ij}F(t^{n,j},z^{n,j})}, & \mbox{  } i = 1,...q, \\{z^{n+1}=z^n + h\sum\limits_{i=1}^{q}w_iF(t^{n,i},z(t^{n,i}))}, & \mbox{  } n = 0,...N-1,
		\end{array}\right.
\end{align}
where every tuple of $\tau_i, w_i, a_{ij}$, is chosen accordingly so that it represent a specific  Runge-Kutta method of $q$ intermediate stages. These number are very often written in form of a matrix, according to the symbolism of Butcher J.C. \cite{butcherBook,butcherCoeff}.
$$\begin{array}{c|c}
	a_{11} \  a_{12} \ \cdots \ a_{1q} & \tau_1 \\ 
	a_{21} \ a_{22} \  \cdots \   a_{2q}  & \tau_2 \\
	\vdots &\vdots \\
	a_{q1} \ a_{q2} \  \cdots \   a_{qq}  & \tau_q \\
	\hline
	b_1 \ b_2 \  \cdots \ b_q \
\end{array}$$
Depending on how we choose $\tau_i, w_i, a_{ij}$ we are able to obtain new Runge-Kutta methods. Specifically, if $a_{ij}=0$ for $j \geq i$ such Runge-Kutta methods are called explicit and can be solved with simple substitutions.  While otherwise they are called implicit and in order to calculate the terms we need to solve a non-linear system for each time-step. For more information see
\cite{butcherBook,butcherRunge,butcherCoeff}.

Also an equivalent and more used form of the system in Eq.~\eqref{104} is 
\begin{align}
	\label{102}
		\left\{\begin{array}{ll}{k^{n,i}=F(t^{n,i},z^{n,i})=F(t^{n,i},y^n + h\sum\limits_{j=1}^{q}a_{ij}k^{n,j})}, & \mbox{ } i=1,...,q,  \\{z^{n+1}=z^n + h\sum\limits_{i=1}^{q}w_ik^{n,i}}, & \mbox{  } n = 0,...N-1,
		\end{array}\right.
\end{align}
where we set $k^{n,i}=F(t^{n,i},z^{n,i})$. This is the form that we are also going to use when we talk about the 4th order Runge-Kutta method.


\begin{thebibliography}{99}
	
	\bibitem{collins}
	G.~W.~Collins II, ``The foundations of celestial mechanics II.'', 1989.
	
	\bibitem{arnold}
	V.~I.~Arnold, ``Mathematical methods of classical mechanics'', 1989.
	
	\bibitem{mayer}
	K.~R.~Meyer and G.~R.~Hall,
	``Introduction to Hamiltonian dynamical systems and the N-body problem'', 1992.
	
	\bibitem{landau}
	L.~D.~Landau and E.~M.~Lifshitz (Trans. by J.~B.~Sykes and J.~S.~Bell),
	``Mechanics: course of theoretical physics'', 2007.
	
	\bibitem{sundman}
	K.~F.~Sundman, Acta Math. {\bf 36} (1913), 105-179.

	\bibitem{poincare}
	H.~Poincaré, ``The three-body problem and the equations of dynamics'', 2017.

	\bibitem{anagnostopoulos}
	K.~N.~Anagnostopoulos, ``Computational physics: a practical introduction to computational physics and scientific computing'', 2014.
	
	\bibitem{ydri}
	B.~Ydri, \href{https://www.worldscientific.com/worldscibooks/10.1142/10283}{doi:10.1142/10283}, 
	\href{https://arxiv.org/abs/1506.02567}{arXiv:1506.02567 [hep-lat]}.

	\bibitem{aarseth}
	S.~J.~Aarseth and F.~Hoyle,
	Astrophysica Norvegica {\bf 9} (1964), 313.
	
	\bibitem{szebehely}
	V.~Szebehely and D.~G.~Bettis,
	Astrophys. Space Sci. {\bf 13} (1971), 365-376.
	
	\bibitem{bettis}
	D.~G.~Bettis, Celes. Mech. {\bf 8} (1973), 229-233.
	
	\bibitem{nacozy}
	P.~E.~Nacozy, Apps {\bf 14} (1971), 40-51.
	
	\bibitem{ahmaad}
	A.~Ahmad and L.~Cohen, Journal of Computational Physics {\bf 12} (1973), 389-402.
	
	\bibitem{zadunoisky}
	 P.~E.~Zadunaisky, Celes. Mech. {\bf 20} (1979), 209-230.
	
	\bibitem{mikkola}
	S.~Mikkola and J.~Hietarinta, Celest. Mech. and Dyn. Astron. {\bf 51} (1991) no.4, 379-394. 
	
	
	\bibitem{musielak}
	Z.~E.~Musielak and B.~Quarles, Reports on Progress in Physics {\bf 77} (2014), 065901, \href{https://arxiv.org/abs/1508.02312}{arXiv:1508.02312 [astro-ph.EP]}.
	
	\bibitem{naoz}
	S.~Naoz, W.~M.~Farr, Y.~Lithwick, F.~A.~Rasio and J.~Teyssandier,
	Mon. Not. Roy. Astron. Soc. \textbf{431} (2013), 2155,
	\href{https://arxiv.org/abs/1107.2414v2}{arXiv:1107.2414 [astro-ph.EP]}.
	
	\bibitem{laves}
	L.~Kurt, Astron. J. {\bf 19} (1898), 97-104.
	
	\bibitem{becker}
	L.~Becker, MNRAS {\bf 80} (1920), 787.
	
	\bibitem{arenstorft}
	R.~F.~Arenstorf,  Celes. Mech. {\bf 14} (1976), 5-9.
	
	\bibitem{wang}
	Q. ~D.~Wang, Acta Astronomica Sinica {\bf 10} (1985), 135-142.
	
	\bibitem{hayli}
	S.~J.~Aarseth, J.~R.~Gott III and E.~L.~Turner, ApJ {\bf 228} (1979), 664-683.
	
	\bibitem{wielen}
	R.~Wielen, Celes. Mech. {\bf 2} (1970), 353-354.
		
	\bibitem{szebehelyStab}
	V.~Szebehely, Celes. Mech. {\bf 22} (1980), 7-12.
	
	\bibitem{scheeres}
	D.~J.~Scheeres, Celest. Mech. Dyn. Astron. {\bf 104} (2009), 103-128.
	
	\bibitem{marchal}
	C.~Marchal, Acta Astronautica {\bf 7}, 555-565.
	
	\bibitem{elmabsoutStab}
	B.~Elmabsout, Academie des Sciences Paris Comptes Rendus Serie B Sciences Physiques {\bf 13} (1989), 1153-1156.
	
	\bibitem{baba}
	L.~K.~Babadzhanyants, Pisma v Astronomicheskii Zhurnal {\bf 7} (1981), 752-755.
	
	\bibitem{dejonghe}
	H.~Dejonghe and P.~Hut, Lecture Notes in Physics {\bf 267} (1986), 212, \href{https://doi.org/10.1007/BFb0116416}{doi:10.1007/BFb0116416}.
		
	\bibitem{Hertz}
	P.~Hertz and  S.~L.~W.~McMillan, Celes. Mech. {\bf 45} (1989), 77.
	
	\bibitem{belleman}
	R.~G.~Belleman, J.~Bedorf and S.~Portegies Zwart,
	New Astron. \textbf{13} (2008), 103-112, 
	\href{https://arxiv.org/abs/0707.0438}{arXiv:0707.0438 [astro-ph]}.
		
	\bibitem{szszodr}
	B.~Szczodrowska, Postepy Astronomii Krakow {\bf 17} (1969), 375-386.
		
	\bibitem{heggie}
	D.~C.~Heggie, Celes. Mech. {\bf 10} (1974), 217-241.	
	
	\bibitem{zare}
	K.~Zare, Celes. Mech. {\bf 10} (1974), 207–15.

	\bibitem{MikkolaReg}
	S.~Mikkola and K.~Tanikawa, Mon. Not. R. Astron. Soc. {\bf 310} (1999), 745–9.
		
	\bibitem{aarsethBOOK}
	S.~J.~Aarseth, Gravitational N-body simulations: tools and algorithms, 2003.
	
	\bibitem{butcherBook}
	J.~C.~Butcher, ``Elementary differential equations and boundary value problems'', 2012.
	
	\bibitem{butcherRunge}
	J.~C.~Butcher, Scholarpedia {\bf 2} (2007), 3147.
	
	\bibitem{freed}
	A.~D.~Freed, \href{https://arxiv.org/abs/1707.02125}{arXiv:1707.02125 [cs.NA]}.
	
	\bibitem{butcherImplicit}
	J.~C.~Butcher, Mathematics of Computation {\bf 18} (1964), 50-64.
	
	\bibitem{butcherStabImplicit}
	J.~C.~Butcher, MBIT Numerical Mathematics {\bf 15} (1964), 358-361.
	
	\bibitem{gautchi}
	W.~Gautschi, ``Numerical Analysis, 2nd edition'', 2012.
		
	\bibitem{butcherCoeff}
	J.~C.~Butcher, Journal of the Australian Mathematical Society {\bf 3} (1963), 185-201.
	
	\bibitem{butcherRungeTree}
	J.~C.~Butcher, Numer Algor {\bf 53} (2010), 153-170.
	
	\bibitem{null}
	G.~W.~Null, W.~M.~Owen and S.~P.~Synnott,
	Astron. J. {\bf 105} (1993), 1993.
	
	\bibitem{pojoman}
	J.~Pojman and V.~Szebehely, ASIC {\bf 246} (1988), 277-288.
	
	
	\end{thebibliography}
\end{document}